\documentstyle[aps,prd]{revtex} 
\begin{document} 
\title{Gravitational vacuum polarization III:\\
Energy conditions in the (1+1) Schwarzschild spacetime\\
gr-qc/9604009} 
\author{Matt Visser\cite{e-mail}} 
\address{Physics Department, Washington University, St. Louis,  
         Missouri 63130-4899} 
\date{1 April 1996} 
\twocolumn[
\maketitle
\parshape=1 0.75in 5.5in
Building on techniques developed in a pair of earlier papers, I
investigate the various point-wise and averaged energy conditions
for the quantum stress-energy tensor corresponding to a
conformally-coupled massless scalar field in the in the (1+1)-dimensional
Schwarzschild spacetime. Because the stress-energy tensors are
analytically known, I can get exact results for the Hartle--Hawking,
Boulware, and Unruh vacua. This exactly solvable model serves as
a useful sanity check on my (3+1)-dimensional investigations wherein
I had to resort to a mixture of analytic approximations and numerical
techniques. Key results in (1+1) dimensions are: (1) NEC is satisfied
outside the event horizon for the Hartle--Hawking vacuum, and
violated for the Boulware and Unruh vacua. (2) DEC is violated
everywhere in the spacetime (for any quantum state, not just the
standard vacuum states).
\vskip 0.125 in
\parshape=1 0.75in 5.5in
PACS number(s): 04.20.-q    04.20.Gz    04.25.-g    04.90.+e
]

\bibliographystyle{unsrt} 
\narrowtext 
\section{INTRODUCTION} 
\def\Implies{\Rightarrow}
 
In a pair of earlier papers~\cite{Visser96a,Visser96b} I have
investigated the gravitational vacuum polarization outside a
Schwarzschild black hole in the Hartle--Hawking and Boulware vacuum
states.

For the Hartle--Hawking vacuum in (3+1) dimensions I found that
the various energy conditions were violated in a  nested set of
onion-like layers located between the event horizon and the unstable
photon orbit~\cite{Visser96a}. Furthermore, based on the Page
approximation, it seems that many of the energy conditions are also
violated inside the event horizon.

For the Boulware vacuum in (3+1) dimensions I found that: (1) All
standard point-wise and averaged energy conditions are violated
throughout the entire region exterior to the event horizon. (2)
Outside the event horizon, all standard point-wise energy conditions
are violated in a maximal manner: they are violated at all points
and for all null/timelike vectors.  (3) Subject to caveats concerning
the applicability and accuracy of the analytic approximation inside
the event horizon, the point-wise energy conditions seem to be
violated even inside the event horizon~\cite{Visser96b}.

In this paper I report a much simpler analysis which serves as a
sanity check on the general formalism. I consider the (1+1)-dimensional
Schwarzschild geometry
\begin{equation}
ds^2 = - (1-2M/r) dt^2 + {dr^2 \over (1-2M/r)}.
\end{equation}
More precisely, I will work with the maximal analytic extension of
this geometry, the (1+1)-dimensional Kruskal--Szekeres manifold.

This geometry is often grandiosely referred to as the (1+1)-dimensional
black hole, and has many interesting features analogous to the
(3+1)-dimensional black hole. One aspect that is very different
from (3+1) dimensions is that the quantum stress energy tensor is
explicitly calculable.

In this paper I will treat all three standard vacuum states:
Hartle--Hawking, Boulware, and Unruh. I also address all point-wise
energy conditions, NEC, WEC, SEC, and DEC, and finally discuss the
ANEC.  The analysis of this paper can also be viewed as an extension
of the (1+1)-dimensional aspects of the recent papers by Ford and
Roman~\cite{Ford-Roman93,Ford-Roman96}.  The results obtained for
the Hartle--Hawking and Boulware vacuum states are qualitatively
similar to the (3+1)-dimensional case~\cite{Visser96a,Visser96b},
and give us additional confidence in the general features deduced
from the numerical analysis required in (3+1) dimensions.

\section{Vacuum polarization in (1+1) Schwarzschild spacetime} 
 
For a static (1+1) dimensional spacetime one knows that 
\begin{equation} 
\langle 0 | T^{\hat\mu}{}_{\hat\nu} | 0 \rangle \equiv  
\left[ \matrix{-\rho&-f\cr 
               +f&-\tau\cr} \right]. 
\end{equation} 
Where $\rho$, $\tau$ and $f$ are functions of $r$, $M$ and $\hbar$.
(Note: I set $G\equiv1$, and choose to work in a local-Lorentz
basis attached to the fiducial static observers [FIDOS].)

A subtlety arises when working in a local-Lorentz basis and looking
at the two-index-down (or two-index-up) versions of the stress-energy.
Outside the horizon one has
\begin{equation} 
g^{\hat\mu\hat\nu}|_{\rm outside} =
\left[ \matrix{-1&0\cr 
               0&+1\cr} \right]. 
\end{equation} 
Consequently
\begin{equation} 
\langle 0 | T^{\hat\mu\hat\nu} | 0 \rangle|_{\rm outside}  =
\left[ \matrix{+\rho&+f\cr 
               +f&-\tau\cr} \right]. 
\end{equation} 
Inside the horizon on the other hand, it is the radial direction
that is timelike, so
\begin{equation}
g^{\hat\mu\hat\nu}|_{\rm inside} = 
\left[ \matrix{+1&0\cr 
               0&-1&\cr} \right]. 
\end{equation} 
Consequently one has the potentially confusing result that
\begin{equation} 
\langle 0 | T^{\hat\mu\hat\nu} | 0 \rangle|_{\rm inside}  =  
\left[ \matrix{-\rho&-f\cr 
               -f&+\tau&\cr} \right]. 
\end{equation} 
Thus, inside the horizon, one should interpret $\tau$ as the energy
density and $\rho$ as the tension (this tension now acting in the
spacelike $t$-direction).

To start the actual analysis I require explicit analytic formulae
for the stress-energy tensor. By working from the analysis in
Christensen and Fulling~\cite[pages 2091--2093]{Fulling77}, or the
presentation in the textbook by Birrell and Davies~\cite[pages
283--285]{Birrell-Davies}, it is easy to show that the stress energy
tensor is given by simple rational polynomial formulae in the variable
$z=(2M/r)$.

In the Hartle--Hawking vacuum:
\begin{eqnarray} 
\rho(z) &=&  
+ p_\infty \; (1 + z + z^2 - 7 z^3), 
\\ 
\tau(z) &=& 
- p_\infty \; (1+z) \; (1+z^2).
\end{eqnarray} 
Here I have defined a constant 
\begin{equation} 
p_\infty \equiv {\hbar\over 6 (16\pi) (2M)^2}.  
\end{equation} 
In the Hartle--Hawking vacuum $p_\infty$ can be interpreted as
the pressure at spatial infinity.

In the Boulware vacuum:
\begin{eqnarray} 
\rho(z) &=&  
-p_\infty \; z^3 \; { 8-7 z \over 1-z }, 
\\ 
\tau(z) &=& 
+ p_\infty \; z^4 {1\over 1-z}.
\end{eqnarray} 

In the Unruh vacuum:
\begin{eqnarray} 
\rho(z) &=&  
+ p_\infty \; { 1 - 16 z^3 + 14 z^4 \over 2 (1-z)}, 
\\ 
\tau(z) &=& 
- p_\infty \; { 1- 2 z^4 \over 2 (1-z) }.
\\ 
f(z) &=& 
+ p_\infty \; { 1  \over 2 (1-z) }.
\end{eqnarray} 

To get these expressions I have written (in the notation of
Christensen and Fulling)~\cite{Fulling77}
\begin{eqnarray}
H_2(z)  &=&  p_\infty \; (1-z^4), \\
T(z)    &=&  8 \; p_\infty \; z^3,
\end{eqnarray}
and carried through the straightforward analysis indicated in
Birrell and Davies~\cite{Birrell-Davies}. For the Boulware and
Unruh vacua the resulting stress-energy tensors have been checked
by taking the explicit formulae of Unruh~\cite{Unruh77}, and Ford
and Roman~\cite{Ford-Roman93,Ford-Roman96}, and translating them
into a local orthonormal basis. With these analytic formulae in
hand, investigation of the energy conditions is straightforward.

I mention in passing several cautionary notes: In (1+1) dimensions one has
\begin{eqnarray} 
&&
\langle U^+ | T^{\hat\mu\hat\nu} | U^+ \rangle +
\langle U^- | T^{\hat\mu\hat\nu} | U^- \rangle = 
\nonumber\\ 
&&\qquad
\langle H | T^{\hat\mu\hat\nu} | H \rangle +
\langle B | T^{\hat\mu\hat\nu} | B \rangle. 
\end{eqnarray} 
Here $U^\pm$ denote the ordinary and time-reversed Unruh vacuum
states.  This special relationship does not survive in (3+1)
dimensions~\cite[page 2098]{Fulling77}. The fact that it happens
to work in (1+1) dimensions is a result of the conformal flatness
of (1+1)-dimensional spacetimes, which implies that all the asymptotic
scattering amplitudes are unity.

A second cautionary note: In (1+1) dimensions, one has the exact result
\begin{equation} 
\langle H | T^{\hat\mu\hat\nu} | H \rangle -
\langle B | T^{\hat\mu\hat\nu} | B \rangle =
p_\infty\; {1\over 1-z} 
\left[ \matrix{+1&0\cr 
               0&-1&\cr} \right]. 
\end{equation} 
Thus the difference between the Hartle--Hawking and Boulware
stress-energy  is exactly a thermal distribution of massless particles
at the Hawking temperature. Despite an early conjecture~\cite[page
2101, equation (6.29)]{Fulling77}, this special relationship does
not survive in (3+1) dimensions~\cite{JLO92}. The fact that it
happens to work in (1+1) dimensions is again a result of the
conformal flatness of (1+1)-dimensional spacetimes.

A final cautionary note is that a subtlety arises when I turn to
discussing the SEC: one must first decide how exactly to continue
the SEC to (1+1) dimensions. In (3+1) dimensions one writes the
SEC as
\begin{equation} 
\bar T_{\mu\nu} V^\mu V^\nu \geq 0? 
\end{equation} 
Where $\bar T$ is the trace-reversed stress tensor
\begin{equation} 
\bar T_{\mu\nu} \equiv   T_{\mu\nu} -  {1\over2}g_{\mu\nu}  T.
\end{equation} 
In (1+1) dimensions one has to decide whether to literally retain
the above definition,
or whether to use the (1+1) dimensional version of trace reversal
\begin{equation} 
\bar T_{\mu\nu} \equiv   T_{\mu\nu} -  g_{\mu\nu}  T.
\end{equation} 

Under the first option, working outside the horizon,
\begin{eqnarray} 
&\langle 0 | \bar T^{\hat\mu\hat\nu} | 0 \rangle& \equiv  
\left[ \matrix{+(\rho-\tau)/2&f\cr 
               f&+(\rho-\tau)/2\cr} \right].
\end{eqnarray}
This option is uninteresting, because with this definition the SEC
is identical to the NEC. To see this note that with this definition
the SEC would be ($\beta\in[0,1]$)
\begin{equation}
\gamma^2 [ (\rho-\tau)/2 \pm 2\beta f + \beta^ 2 (\rho-\tau)/2 ] \geq 0?
\end{equation}
This is easily rearranged to give
\begin{equation}
(\rho-\tau) \pm 4{ \beta \over 1+\beta^2} f \geq 0?
\end{equation}
Since this is to hold for all $\beta\in[0,1]$, this implies {\em
and is implied by}
\begin{equation}
(\rho-\tau) \pm 2 f \geq 0?
\end{equation}
Which is exactly the NEC.

On the other hand, with the second option one has
\begin{equation}
\langle 0 | \bar T^{\hat\mu\hat\nu} | 0 \rangle \equiv  
\left[ \matrix{-\tau&f\cr 
               f&+\rho\cr} \right]. 
\end{equation}
With this definition the SEC is
\begin{equation}
\gamma^2 [ -\tau \pm 2\beta f + \beta^ 2 \rho ] \geq 0?
\end{equation}
This is equivalent to
\begin{equation}
-\tau \geq 0?  \qquad \rho-\tau \pm 2 f \geq 0?
\end{equation}
Thus this definition of SEC implies {\em but is not implied by} the NEC.

\section{Hartle--Hawking vacuum} 
\subsection{Outside the horizon:} 
 
Outside the event horizon, the NEC reduces to the single constraint 
\begin{equation} 
\rho(r) - \tau(r) \geq 0? 
\end{equation} 
It is easy to see that
\begin{equation} 
\rho(r) - \tau(r) = 2 \; p_\infty \; (1 - z) \; (1 + 2 z + 3 z ). 
\end{equation} 
This is explicitly positive everywhere outside the event horizon.
Therefore the NEC is definitely satisfied everywhere outside the
event horizon.

Outside the event horizon, the WEC reduces to the pair of constraints 
\begin{equation} 
\rho \geq 0?    \qquad 
\rho(r) - \tau(r) \geq 0?
\end{equation} 
It is easy to see that $\rho$ switches sign and becomes negative
for $z>0.671907$ corresponding to $r<2.9776 M$.  (One has to
numerically solve a cubic.) Therefore the WEC is definitely violated
in the region $r\in[2M,2.9776 M]$.

Outside the event horizon, the DEC reduces to the three constraints 
\begin{equation} 
\rho(r) \geq 0? 
\qquad \rho(r) - \tau(r) \geq 0?
\qquad \rho(r) + \tau(r) \geq 0? 
\end{equation} 
It is easy to see that
\begin{equation} 
\rho(r) + \tau(r) = -8 \; p_\infty \; z^3. 
\end{equation} 
(This is in fact just the negative of the anomalous trace.) This
is explicitly negative everywhere outside the event horizon.
Therefore the DEC is definitely violated everywhere outside the
event horizon.

Outside the event horizon,  the SEC reduces to the pair of constraints 
\begin{equation} 
-\tau(r) \geq 0? 
\qquad
\rho(r) -\tau(r)\geq 0?
\end{equation} 
Both of these quantities are positive outside the horizon, so SEC
is satisfied everywhere outside the horizon.

\subsection{Inside the horizon:} 
 
Inside the event horizon, the radial coordinate becomes timelike,
and the roles played by $\rho(r)$ and $\tau(r)$ are interchanged.
The NEC reduces to the constraint
\begin{equation} 
\tau(r) -\rho(r)\geq 0?  
\end{equation} 
We have already seen that $\rho-\tau$ explicitly factorizes, goes
to zero and switches sign at the event horizon. This sign switch
in $\rho-\tau$ exactly matches the definition switch in the NEC.
We conclude that the NEC is definitely satisfied everywhere inside
the event horizon. 

Inside the event horizon, the WEC reduces to the pair of constraints 
\begin{equation} 
\tau(r)  \geq 0?    \qquad 
\tau(r) - \rho(r)\geq 0?
\end{equation} 
But $\tau$ is negative for all $r$.  Therefore the WEC is definitely
violated inside the event horizon. This automatically implies that
the DEC is definitely violated inside the event horizon.

Inside the event horizon,  the SEC reduces to the pair of constraints 
\begin{equation} 
-\rho(r) \geq 0? 
\qquad
\tau(r)-\rho(r)\geq 0?
\end{equation} 
Both of these quantities are positive inside the horizon, so the
SEC  is satisfied everywhere inside the horizon.

\subsection{Summary:} 

In the Hartle--Hawking vacuum:
\begin{itemize}
\item
NEC and SEC are satisfied throughout the spacetime.
\item
WEC is satisfied for $r\in[2.9776 M,\infty]$ and violated for
$r\in[0,2.9776 M]$
\item
DEC is violated throughout the spacetime.
\end{itemize}

\section{Boulware vacuum} 
\subsection{Outside the horizon:} 
 
Outside the event horizon, the NEC reduces to the single constraint 
\begin{equation} 
\rho(r) - \tau(r) \geq 0? 
\end{equation} 
It is easy to see that
\begin{equation} 
\rho(r) - \tau(r) = - 2 \; p_\infty \; {4 - 3z \over 1 - z} . 
\end{equation} 
This is explicitly negative everywhere outside the event horizon.
Therefore the NEC, (and also WEC, DEC, and SEC) are definitely violated
everywhere outside the event horizon.

\subsection{Inside the horizon:} 
 
Inside the event horizon, we interchange  $\rho(r)$ and $\tau(r)$.
The NEC reduces to the constraint
\begin{equation} 
\tau(r) -\rho(r)\geq 0?  
\end{equation} 
For the Boulware vacuum, $\tau-\rho$ is positive in the range
$z\in[0,1]\cup[4/3,\infty]$. Thus NEC is violated in the range
$r\in[3M/2, 2M]$ and satisfied in the range $r\in[0, 3M/2]$.

Inside the event horizon, the WEC reduces to the pair of constraints 
\begin{equation} 
\tau(r)  \geq 0?    \qquad 
\tau(r) - \rho(r)\geq 0?
\end{equation} 
But $\tau$ is negative inside the horizon.   Therefore the WEC (and
also the DEC) is definitely violated everywhere inside the event horizon.

Inside the event horizon,  the SEC reduces to the pair of constraints 
\begin{equation} 
-\rho(r) \geq 0? 
\qquad
\tau(r)-\rho(r)\geq 0?
\end{equation} 
For the Boulware vacuum, $-\rho$ is positive in the range
$z\in[0,1]\cup[8/7,\infty]$, while $\tau-\rho$ is positive in the
range $z\in[0,1]\cup[4/3,\infty]$. Thus SEC is again violated
in the same range as the NEC.

\subsection{Summary:} 

In the Boulware vacuum:
\begin{itemize}
\item
NEC and SEC are violated for $r\in[3M/2,\infty]$ and
satisfied for $r\in[0,3M/2]$ .
\item
WEC  and DEC are violated throughout the spacetime.
\end{itemize}

\section{Unruh vacuum} 
\subsection{Outside the horizon:} 
 
Outside the event horizon, the presence of a flux in the Unruh
vacuum implies that the NEC reduces to a pair of constraints
\begin{equation} 
\rho(r) - \tau(r) \pm  2 f(r)\geq 0? 
\end{equation} 
(Warning: The minus sign corresponds to an outgoing null ray, while
the plus sign represents an ingoing null ray.) It is easy to see
that
\begin{eqnarray} 
\rho(r) - \tau(r) + 2 f(r) &=& 2 \; p_\infty \; (1-z) (1+ 2z + 3 z^2) . 
\\ 
\rho(r) - \tau(r) - 2 f(r) &=& - 2 \; p_\infty z^3 \; {4-3z\over 1-z}. 
\end{eqnarray}
The second of these constraints is explicitly negative everywhere
outside the event horizon.  Therefore the NEC, (and also the WEC,
DEC, and SEC) are definitely violated everywhere outside
the event horizon.

You may wish to note that
\begin{eqnarray} 
[\rho(r) - \tau(r) + 2 f(r)]|_{\rm U}  &=& [\rho(r) - \tau(r)]|_{\rm H}, 
\\ 
\null
[\rho(r) - \tau(r) - 2 f(r)]|_{\rm U}  &=& [\rho(r) - \tau(r)]|_{\rm B}. 
\end{eqnarray}
Thus the discussion for the Hartle--Hawking and Boulware vacua can
be carried over immediately to respectively the ingoing and outgoing
null geodesics of the Unruh vacuum. [This result is special to
(1+1) dimensions.]

A subtlety is that it is now possible to define two distinct types
of NEC, a NEC$^+$ and a NEC$^-$, depending on whether one
wishes to follow outgoing or ingoing null curves. (That is,
depending on whether one is approaching Scri$^+$ or coming in from
Scri$^-$.)

The NEC$^+$ condition (outgoing null curves) is violated outside
the event horizon, while the NEC$^-$ condition (ingoing null curves)
is satisfied.

Notice that it does not make sense to talk about WEC$^\pm$, DEC$^\pm$
or SEC$^\pm$ because the use of timelike vectors in these energy
conditions does not let you make an invariant separation into
classes of ingoing and outgoing geodesics.

\subsection{Inside the horizon:} 
 
Inside the event horizon, we again interchange  $\rho(r)$ and $\tau(r)$.
The NEC reduces to the pair of constraints
\begin{equation} 
\tau(r) -\rho(r) \mp 2f(r) \geq 0?  
\end{equation} 
These are identical to the two conditions outside the horizon with
the signs flipped.

In particular the NEC$^-$ condition is satisfied inside the event
horizon, while the NEC$^+$ condition is violated for $z\in[1,4/3]$
and satisfied for $z\in[4/3,\infty]$, corresponding to $r\in[3M/2,2M]$
and $r\in[0, 3M/2]$ respectively.

This is enough to tell us that WEC, DEC, and SEC are violated at
least in the region $r\in[3M/2,2M]$.

Inside the event horizon, the WEC is
\begin{equation} 
\gamma^2(\tau \mp2 \beta f -\beta^2 \rho) \geq 0?
\end{equation} 
which reduces to the pair of constraints 
\begin{equation} 
\tau(r)  \geq 0?    \qquad 
\tau(r) - \rho(r) \pm 2 f(r) \geq 0?
\end{equation} 
But $\tau$ is negative inside the horizon.   Therefore WEC (and
also DEC) is definitely violated everywhere inside the event horizon.

Inside the event horizon the  SEC is
\begin{equation} 
\gamma^2(-\rho\mp2\beta f +\beta^2 \tau) \geq 0?
\end{equation} 
which reduces to the triplet of constraints 
\begin{equation} 
-\rho(r) \geq 0? 
\qquad
\tau(r)-\rho(r) \pm 2 f\geq 0?
\end{equation} 
For the Unruh vacuum, $\rho$ is positive in the range $z\in[1,1.08729]$,
(and $r\in[0,0.474574]$), while the second condition  is  always
satisfied, and the third condition is violated in the range
$z\in[0,1]\cup[4/3,\infty]$. Thus SEC is again violated in the
same range as the NEC.

\subsection{Summary:} 

In the Unruh vacuum:
\begin{itemize}
\item
NEC and SEC are violated for $r\in[3M/2,\infty]$ and
satisfied for $r\in[0,3M/2]$ .
\item 
NEC$^-$ is satisfied throughout the spacetime.
\item
NEC$^+$ is  violated for $r\in[3M/2,\infty]$ and
satisfied for $r\in[0,3M/2]$ .
\item
WEC  and DEC are violated throughout the spacetime.
\end{itemize}

\section{Total DEC violation} 

I shall now show that in any quantum state, vacuum or not, the DEC
is violated throughout the spacetime. (This comes from the fact that
DEC violations in 1+1 dimensions can be intimately related to the
trace anomaly.)

I start from the fact that for the DEC to hold, the vector
$T^{\mu\nu}V_\nu$ must be non-spacelike for any timelike vector
$V_\nu$. This requires that for all $\beta\in [-1,1]$ one must have
\begin{equation}
| f - \beta\tau | \leq | \rho + \beta f | ?
\end{equation}
Thus
\begin{eqnarray}
&&
(f - \beta\tau)^2 \leq (\rho + \beta f )^2  ?
\nonumber\\
&&
\Implies f^2 \mp 2 |\beta| f \tau + \beta^2 \tau^2 \leq 
\rho^2 \pm 2 |\beta| f \rho + \beta^2 f^2 ?
\nonumber\\
&&
\Implies f^2 + \beta^2 \tau^2 \leq  \rho^2 + \beta^2 f^2 ?
\nonumber\\
&&
\Implies f^2 + \tau^2 \leq  \rho^2 +  f^2 ?
\nonumber\\
&&
\Implies \tau^2 \leq  \rho^2 ?
\nonumber\\
&&
\Implies \rho \pm \tau \geq 0 ?
\end{eqnarray}
Thus a minimum condition for DEC to hold is for $\rho+\tau = -\langle
T \rangle$ to be positive. But we know that $\langle T \rangle$ is
given exactly by the conformal anomaly (and this is independent of
the quantum state), and that in the 1+1 Schwarzschild geometry
$\langle T \rangle =  8\; p_\infty \; z^3$ is positive. Thus DEC
is violated everywhere in the spacetime.

Of course this result generalizes to any (1+1)-dimensional spacetime:
We now know that the DEC must be violated at least on those regions
where
\begin{equation}
\langle T \rangle\equiv {1\over 768 \pi } R > 0.
\end{equation}
That is: one needs a negative Ricci scalar to even have a hope of
satisfying the DEC.

\section{ANEC violation?} 
 
Outside the event horizon we may certainly write the ANEC integral
as~\cite{Visser96a} (See also~\cite[page 133, equations
(12.59)--(12.63)]{Visser}.)
\begin{eqnarray}
I_\gamma 
&\equiv&
\int_\gamma T_{\mu\nu} \; k^\mu \; k^\nu \; d\lambda,
\nonumber\\
&=&
\int_\gamma (\rho -\tau \pm 2 f) \; \xi^2 \; d\lambda,
\nonumber\\
&=&
\int_\gamma (\rho -\tau \pm 2 f) \;  dt,
\nonumber\\
&=&
\int_{2M}^{\infty} {(\rho -\tau \pm 2 f) \over (1-2M/r)}\; dr.
\end{eqnarray}
This appears to weight the region near the event horizon very
heavily---because of the explicit pole at $r=2M$.  However, the
integrand $(\rho-\tau \pm 2f)$ often has a zero at the event horizon.
(This occurs in the Hartle--Hawking state, and for the ingoing null
geodesics in the Unruh state).

Inside the event horizon there are additional sign-flips:
\begin{eqnarray}
I_\gamma 
&\equiv&
\int_\gamma T_{\mu\nu} \; k^\mu \; k^\nu \; d\lambda,
\nonumber\\
&=&
\int_\gamma (\tau -\rho \mp 2 f) \; \xi^2 \; d\lambda,
\nonumber\\
&=&
\int_\gamma (\tau -\rho \mp 2 f)  \;  dt,
\nonumber\\
&=&
-\int_\gamma {(\tau -\rho \mp 2 f) \over (1-2M/r)}\; dr.
\nonumber\\
&=&
\int_0^{2M} {(\rho -\tau \pm 2 f) \over (1-2M/r)}\; dr.
\end{eqnarray}
To see where the extra minus sign comes, assume we are looking at
an outgoing null geodesic (one that approaches Scri$^+$), such a
null geodesic starts off from the past singularity and must first
pass through the past event horizon H$^-$. Inside the past event
horizon,  if you want an outgoing null geodesic (increasing $r$)
to be travelling forward in $t$ one must take $dr = |1-2M/r|dt$.
Reversing the argument, the same result holds for incoming null
geodesics, ones that start out from Scri$^-$, cross the future
event horizon H$^+$, and terminate on the future singularity.  (The
present discussion does not address null geodesics that pass through
the bifurcation two-point, see Ford and Roman~\cite{Ford-Roman96}.)

Combining these results, one may formally write
\begin{eqnarray}
I_\gamma
&=& \int_{0}^{\infty} 
{(\rho -\tau \pm 2 f) \over (1-2M/r)}\; dr
\nonumber\\
&=& 2 M \int_{0}^{\infty} 
{(\rho -\tau \pm 2 f) \over z^2(1-z)}\; dz. 
\end{eqnarray}
The integral can potentially have divergences at $r=0$, $r=2M$,
and $r=\infty$.

For the Hartle--Hawking vacuum everything is well-defined at the
event horizon. It is most sensible to integrate outward or inward
from $r=2M$, to consider
\begin{eqnarray}
I[z] 
&=&  2 M \int_{z}^{1} 
{(\rho -\tau) \over \bar z^2(1-\bar z)}\; d\bar z
\nonumber\\
&=& 4 M p_\infty \{ (1-z)(1+3z)/z - 2\ln(z) \} 
\end{eqnarray}
Note that for $z<1$ (that is, $r>2M$) this is explicitly positive
(as it should be). There is of course an infra-red divergence as
$r\to \infty$.  For $z>1$ one should switch the integration limits,
and again get a positive result. There is now an ultraviolet
singularity as $z\to \infty$ ($r\to 0$). 

The total ANEC integral, from $r=\infty$ to $r=0$, is positive infinity.

For the Boulware vacuum there is a singularity at the horizon, so
it is more sensible to integrate inward from spatial infinity and
keep $z<1$:
\begin{eqnarray}
I[z] 
&=&  2 M \int_{0}^{z} 
{(\rho -\tau) \over \bar z^2(1-\bar z)}\; d\bar z
\nonumber\\
&=& 4 M p_\infty  \{ z(2-3z)/(1-z)  +2 \ln(1-z) \} 
\end{eqnarray}
Although the polynomial piece here is (for $z<2/3$) positive it is
easy enough to check that the logarithm is negative and dominant.
If one tries to push this integral past the event horizon one picks
up a negative infinity.

Inside the event horizon one has singularities at both $r=0$ and
$r=2M$. It is perhaps most instructive to fix one end of the ANEC
integral at $z=4/3$, the boundary between the NEC satisfying and
NEC violating region. In that case
\begin{eqnarray}
I[z] 
&=&  2 M \int_{4/3}^{z} 
{(\rho -\tau) \over \bar z^2(1-\bar z)}\; d\bar z
\nonumber\\
&=&  4 M p_\infty  
\left\{  
{(z-2)(3z-4) \over (z-1) } + 
2 \ln[3(z-1)] 
\right\} 
\end{eqnarray}
For $z> 4/3$: Although the polynomial piece here changes sign at
$z=2$, it is easy enough to check that the logarithm is positive
and always sufficient to make the net integral positive. Consequently,
integrating inward from $r=3M/2$ toward $r=0$ gives a positive
contribution to the ANEC.\\
For $z<4/3$: The polynomial piece is positive and  the logarithm
is negative and sub-dominant. Thus, after switching the limits of
integration, one sees that integrating inward from just inside
$r=2M$ to $r=3M/2$ gives a negative contribution to the ANEC.

If one insists on integrating all the way from $r=\infty$ to $r=0$
one picks up two negative infinities and one positive infinity---for
a net result that is at best purely formal. (With suitable regulator,
one might argue that the overall integral can be taken to be negative
infinity.)  This is not in conflict with the ANEC theorems of
Yurtsever~\cite{Yurtsever90a}, and Wald and
Yurtsever~\cite{Wald-Yurtsever}, because those theorems were derived
in asymptotically flat (1+1)-dimensional spacetimes with technical
assumptions about the inextendible nature of the null geodesics.
The singularity present in the Schwarzschild geometry prevents us
from applying these theorems to the present case.

For ingoing null geodesics in the Unruh vacuum, the discussion is
identical to that for the Hartle--Hawking vacuum, while for outgoing
null geodesics in the  in the Unruh vacuum, the discussion is
identical to that for the Boulware vacuum.

\section{Discussion} 
 
In a pair of companion papers~\cite{Visser96a,Visser96b}, I have
studied the gravitational vacuum polarization in (3+1) dimensions,
discussing both the Hartle--Hawking and Boulware vacuum states. In
the (3+1) Hartle--Hawking vacuum state I discovered  a complicated
layering of energy-condition violations confined to the region
between the unstable photon orbit and the event horizon.  In the
(3+1) Boulware vacuum I found that all point-wise energy and averaged
conditions are violated throughout the entire region exterior to
the event horizon, and that the point-wise energy conditions seemed
to be violated inside the event horizon.

In this paper I have looked at the analytically more tractable
model of (1+1)-dimensional Schwarzschild spacetime, mainly as a
sanity check on the (3+1)-dimensional calculations. Happily, the
basic flavour of the (3+1)-dimensional results follows through:
Many of the energy conditions are violated, the precise locations
and manner of violation being influenced by the particular vacuum
state being considered. The results of this paper are also compatible
with, and extensions of, earlier (1+1)--dimensional investigations
by Ford and Roman~\cite{Ford-Roman93,Ford-Roman96}.

The ANEC is satisfied for the Hartle--Hawking vacuum, but is
ill-defined (and arguably negative) for the Boulware vacuum.
(Certainly integrating from infinity down to the event horizon
gives a infinite and negative ANEC integral.)

Overall the situation is this: energy condition violations are
ubiquitous and are particularly prevalent in the Boulware vacuum
in the region outside the event horizon, where every point-wise
energy condition is violated, and all one-sided integrated energy
conditions are violated. (The same comment applies to outgoing null
curves in the Unruh vacuum.)

Continuing the discussion back to (3+1) dimensions: These ubiquitous
violations of the energy conditions have immediate and significant
impact on issues such as whether or not it is possible, even in
principle, to generalise the classical singularity
theorems~\cite{Hawking-Ellis}, classical positive mass
theorems~\cite{Penrose-Sorkin-Woolgar}, and classical laws of black
hole dynamics to semiclassical quantum gravity~\cite{Visser95,Visser}.
It seems that the standard energy conditions may not be the right
tools for the job. Something along the lines of the Ford--Roman
``quantum inequalities'' might be more
useful~\cite{Visser96b,Ford-Roman94,Ford-Roman95b}.

\acknowledgements

I wish to thank Nils Andersson, \'Eanna Flanagan, Larry Ford, and
Tom Roman for their comments and advice.

The numerical analysis in this paper was carried out with the aid
of the Mathematica symbolic manipulation package.

This research was supported by the U.S. Department of Energy.
 
 

\begin{references} 
\bibitem[*]{e-mail}Electronic mail: visser@kiwi.wustl.edu 
\begin{thebibliography}{10}

\bibitem{Visser96a}
M.~Visser.
\newblock Gravitational vacuum polarization I:  Energy conditions in the
  {Hartle--Hawking} vacuum.
\newblock {\em gr-qc/9604007}, 1996.

\bibitem{Visser96b}
M.~Visser.
\newblock Gravitational vacuum polarization II:  Energy conditions in the
  {Boulware} vacuum.
\newblock {\em gr-qc/9604008}, 1996.

\bibitem{Ford-Roman93}
L.~H. Ford and T.~A. Roman.
\newblock Motion of inertial observers through negative energy.
\newblock {\em Phys. Rev. D}, 48:776--782, 1993.

\bibitem{Ford-Roman96}
L.~H. Ford and T.~A. Roman.
\newblock Averaged energy conditions and evaporating black holes.
\newblock {\em Phys. Rev. D}, 53:1988--2000, 1996.

\bibitem{Fulling77}
S.~M. Christensen and S.~A. Fulling.
\newblock Trace anomalies and the {Hawking} effect.
\newblock {\em Phys. Rev. D}, 15:2088--2104, 1977.

\bibitem{Birrell-Davies}
N.~D. Birrell and P.~C.~W. Davies.
\newblock {\em Quantum Fields in Curved Space}.
\newblock Cambridge University Press, Cambridge, England, 1982.

\bibitem{Unruh77}
W.~D. Unruh.
\newblock Origin of the particles in black-hole evaporation.
\newblock {\em Phys. Rev. D}, 15:365--369, 1977.

\bibitem{JLO92}
B.~P. Jensen, J.~G. McLauglin, and A.~C. Ottewill.
\newblock Anisotropy of the quantum thermal state in {Schwarzschild} spacetime.
\newblock {\em Phys. Rev. D}, 45:3002--3005, 1992.

\bibitem{Visser}
M.~Visser.
\newblock {\em Lorentzian wormholes --- from Einstein to Hawking}.
\newblock American Institute of Physics, New York, 1995.

\bibitem{Yurtsever90a}
U.~Yurtsever.
\newblock Does quantum field theory enforce the averaged weak energy condition?
\newblock {\em Class. Quantum Grav.}, 7:L251--L258, 1990.

\bibitem{Wald-Yurtsever}
R.~M. Wald and U.~Yurtsever.
\newblock General proof of the averaged null energy condition for a massless
  scalar field in two dimensional spacetime.
\newblock {\em Phys. Rev. D}, 44:403--416, 1991.

\bibitem{Hawking-Ellis}
S.~W. Hawking and G.~F.~R. Ellis.
\newblock {\em The Large Scale Structure of Space-Time}.
\newblock Cambridge University Press, Cambridge, England, 1973.

\bibitem{Penrose-Sorkin-Woolgar}
R.~Penrose, R.~D. Sorkin, and E.~Woolgar.
\newblock A positive mass theorem based on the focusing and retardation of null
  geodesics.
\newblock 1993.
\newblock gr-qc/9301015.

\bibitem{Visser95}
M.~Visser.
\newblock Scale anomalies imply violation of the averaged null energy
  condition.
\newblock {\em Phys. Lett.}, B349:443--447, 1995.

\bibitem{Ford-Roman94}
L.~H. Ford and T.~A. Roman.
\newblock Averaged energy conditions and quantum inequalities.
\newblock {\em Phys. Rev. D}, 51:4277--4286, 1995.

\bibitem{Ford-Roman95b}
L.~H. Ford and T.~A. Roman.
\newblock Averaged energy conditions constrain traversable wormhole geometries.
\newblock {\em gr-qc/9510071}, 1995.

\end{thebibliography}
\end{references}
\end{document}